\documentclass[aps,pre,showpacs,superscriptaddress]{revtex4}
\usepackage{ifpdf}
\ifpdf
\usepackage[pdftex]{graphicx}
\DeclareGraphicsExtensions{.pdf,.PDF} 
\usepackage{hyperref}
\else
\usepackage[dvips]{graphicx}
\DeclareGraphicsExtensions{.eps,.EPS}
\fi
\usepackage{color,cancel}

\begin{document}
\title{Brownian walkers within subdiffusing territorial boundaries}

\author{L. Giuggioli}
\affiliation{Bristol Centre for Complexity Sciences, University of Bristol, Bristol, UK}
\affiliation{Department of Engineering Mathematics, University of Bristol, Bristol, UK}
\affiliation{School of Biological Sciences, University of Bristol, Bristol, UK}

\author{J. R. Potts}
\affiliation{Bristol Centre for Complexity Sciences, University of Bristol, Bristol, UK}
\affiliation{School of Biological Sciences, University of Bristol, Bristol, UK}

\author{S. Harris}
\affiliation{School of Biological Sciences, University of Bristol, Bristol, UK}

\begin{abstract}
Inspired by the collective phenomenon of territorial emergence, whereby animals move and interact through the scent marks they deposit, we study the dynamics of a 1D Brownian walker in a random environment consisting of confining boundaries that are themselves diffusing anomalously. We show how to reduce, in certain parameter regimes, the non-Markovian, many-body problem of territoriality to the analytically tractable one-body problem studied here. The mean square displacement (MSD) of the 1D Brownian walker within subdiffusing boundaries is calculated exactly and generalizes well known results when the boundaries are immobile. Furthermore, under certain conditions, if the boundary dynamics are strongly subdiffusive, we show the appearance of an interesting non-monotonicity in the time dependence of the MSD, giving rise to transient negative diffusion.
\end{abstract}
\maketitle

\section{Introduction and motivation}
\label{sec1}

How disorder affects the intrinsic diffusion of a particle is a question of extreme practical relevance. It is often encountered in transport processes \cite{dresden1961} whenever one attempts to confront theoretical predictions with actual experiments. Not surprisingly, the study of disorder has attracted a lot of interest in a variety of contexts and often with the aim of understanding the validity of the theoretical assumptions of idealized systems \cite{alexanderetal1981,hauskher1987,stanleyostrowsky1993,maykuhnbook2000}. Although many different kinds exist, disordered systems are broadly divided into two classes depending on the type of heterogeneity: static or dynamic \cite{kenkrebook1982}.
When the heterogeneity is due to the presence of some `quenched' observables that do not change over time, one talks about a system with static disorder (see e.g. \cite{hughesbookvol2}). On the other hand, when the heterogeneity emerges as the system evolves in time, one talks about dynamic disorder (see e.g. \cite{bishopetalbook1998}). Of interest here is one of the simplest dynamically disordered system: a random walker constrained to move within an area whose extent is randomly fluctuating.

A problem where dynamic disorder plays a fundamental role is the formation of territorial patterns in the animal kingdom. A recent study \cite{giuggiolietal2011} shows how territories (spatial inhomogeneity) collectively emerge from the spatio-temporal trajectories of individual animals with interactions between them mediated via scent marks. In that study a microscopic stochastic model of territorial random walkers is introduced whereby each animal performs a random walk \cite{okubolevinbook2002} with the added ingredient that, when an animal visits a site, it leaves its scent that lasts for a fixed time $T_{AS}$. During this time, no other animal can visit this site. As a result, each animal has a territory at any given instant $t$. This territory is the set of sites visited by the animal in the time period $[t-T_{AS},t]$. An animal can diffuse only over sites that belong to its own territory and those sites which do not belong to the territory of any other animal. Thus, both the animals and their respective territories move with time, but their growth rates are different. The centroids of the territories subdiffuse, due to the exclusion principle \cite{harris1965,landim1992,schonherrschutz2004}, while the animals themselves move diffusively.

One of our aims is to reduce, in the 1D case, the many-body non-Markovian problem of territory formation to the one-body problem of a random walker moving within a random environment. The wildly different time scales over which territories and animals displace, allows us to find the appropriate one-body effective description. If we focus on a specific individual, we can view it as being subject to reflecting boundary conditions, located at the position of the left and right neighbouring territorial borders, that are changing gradually relative to the time scale over which the walker diffuses. This gradual change in the external conditions (environment) is what characterizes an adiabatic process. We thus invoke an adiabatic approximation, which consists of taking the probability distribution of the walker to be identical to the one governed by a diffusion equation within an interval equal to the instantaneous separation distance between the neighboring territorial borders. The joint probability distribution $P(x,L_{1},L_{2},t)$ of the walker position $x$ and, respectively, the left and right boundary locations $L_{1}$ and $L_{2}$ can be written as $P(x,L_{1},L_{2},t)\approx Q(L_{1},L_{2},t)W(x,t|L_{1},L_{2})$, with $W(x,t|L_{1},L_{2})$ being the walker probability distribution for a given value of $L_{1}$ and $L_{2}$, and with $Q(L_{1},L_{2},t)$, the joint distribution of the two boundary positions, to be determined. Our goal here is to show that it is possible to capture the dynamics of the random walkers in 
the territoriality problem by representing the dynamics of $Q(L_{1},L_{2},t)$ through a Fokker-Planck formalism \cite{riskenbook1989} with time-dependent diffusion coefficients.

The paper is organized as follows. The analytic expression for the probability $P(x,L_{1},L_{2},t)$ of the simplified one-body problem is computed in Sec. \ref{sec2}. The results in Sec. \ref{sec2} are compared, in Sec. \ref{sec3}, with stochastic simulations of two territorial random walkers in an interval with periodic boundary conditions. In Sec. \ref{negativediff} the analytic computation of the MSD is developed to show the appearance of transient negative diffusivity under certain conditions. The paper ends with a brief discussion in Sec. \ref{sec5}.

\section{Joint probability distribution of walker and boundaries}
\label{sec2}
Since territorial boundaries undergo an exclusion process, the probability distribution of the separation distance between the left and right boundaries can be represented at long times by a Gaussian whose width is increasing proportionally to $\sqrt{t}$ rather than $t$, or by a diffusion equation with an appropriate time-dependent diffusion constant \cite{batchelor52,mandelbrotvanness68}. The mean of this Gaussian is centered around the average territory size, which is simply the inverse of the population density \cite{giuggiolietal2011}. In our reduced one-body problem we mimic the presence of a mean territory size and fluctuations around this mean by assuming that the left and right boundaries are attached by a spring, whose rest position equals the mean territory size $L$. This leads us to the use of a Fokker-Planck formalism with a time-dependent diffusion constant  \cite{giuggiolietal2007} and with a constraining quadratic potential $U(L_{1},L_{2})=\gamma(t)(L_{2}-L_{1}-L)^{2}/4$ of the form
\begin{equation}
\frac{\partial Q(L_{1},L_{2},t)}{\partial t}=K\varphi(t)\left(\frac{\partial^{2}}{\partial L_{1}^2}+\frac{\partial^{2}}{\partial L_{2}^2}\right)Q(L_{1},L_{2},t)+\frac{\gamma(t)}{2}\left(\frac{\partial}{\partial L_{2}}-\frac{\partial}{\partial L_{1}}\right)\left[(L_{2}-L_{1}-L)Q(L_{1},L_{2},t)\right],
\label{fptdc}
\end{equation}
The 2D force that $U(L_{1},L_{2})$ generates, that is $\gamma(t)(L_{2}-L_{1}-L)/2$ along $L_{1}$ and $-\gamma(t)(L_{2}-L_{1}-L)/2$ along $L_{2}$, has the effect of making the mean distance between the two boundaries equal $L$ at long times. Counter to this driving force is the boundary spreading due to the random fluctuations associated with the positive function $\varphi(t)$. For the time dependence of $\gamma(t)$ we focus on the case in which $\gamma(t)=\gamma\varphi(t)$ and to the constant case, i.e. when $\gamma(t)=\gamma$. In the former case we talk about a subordinated stochastic process since Eq. (\ref{fptdc}) can be rewritten via the transformation $\tau=\int_{0}^{t}ds\,\varphi(s)$ as the evolution over the time $\tau$ of a Fokker-Planck equation without time-dependent coefficients. In the latter case we have a time scale disparity between the diffusive process and the time it takes for the boundaries to reach their asymptotic separation value $L$. This, we will show in Sec. \ref{negativediff}, has the effect of generating negative diffusivity under certain conditions.

Through the variable transformation $\lambda=L_{2}-L_{1}$ and $\mathcal{L}=(L_{2}+L_{1})/2$, Eq. (\ref{fptdc}) can be decoupled, i.e. $Q(L_{1},L_{2},t)=\mathcal{Q}_{1}(\mathcal{L},t)\mathcal{Q}_{2}(\lambda,t)$, into a differential equation for the boundary separation $\lambda$ and a differential equation for the boundary centroid location $\mathcal{L}$. These two equations, together with the condition that, at all times, the two boundaries cannot cross each other, i.e. $\vec{\nabla} Q(L_{1},L_{2},t)\cdot \hat{n}=0$ where $\hat{n}$ is the normal to the line of points where $L_{1}=L_{2}$ \cite{ambjornssonetal2008}, can be solved exactly (see details in Appendix A). For localized initial conditions of Dirac $\delta$ type of the form $Q(L_{1},L_{2},0)=\delta(L_{1}+L/2)\delta(L_{2}-L/2)$ with $L_{2,0}=L/2=-L_{1,0}$, the time-dependent solution of Eq. (\ref{fptdc}) with the mentioned boundary condition is given by
\begin{equation}
Q(\lambda,\mathcal{L},t)=H(\lambda)\frac{e^{-\frac{(\lambda-L)^{2}}{b(t)}}+e^{-\frac{(\lambda+L)^{2}}{b(t)}}}{\sqrt{\pi b(t)}}\frac{e^{-\frac{\mathcal{L}^{2}}{c(t)}}}{\sqrt{\pi c(t)}},
\label{QL1L2}
\end{equation}
where $b(t)=8K\int_0^t ds \varphi(s)e^{-2\left(G(t)-G(s)\right)}$ with $G(t)=\int_{0}^{t}ds\,\gamma(s)$, $c(t)=2K\int_{0}^{t}\varphi(s)ds$ and the Heaviside function $H(y)$ is such that $H(y)=1$ ($H(y)=0$) if $y>0$ (if $y<0$). From the form of Eq. (\ref{QL1L2}) it is evident that $b(t)$ controls the diffusion of the boundary separation around the mean value $L$, and $c(t)$ regulates the diffusion of the boundary centroid. For the choice $\gamma(t)=\gamma\varphi(t)$ we have $b(t)=4(K/\gamma)\left\{1-\exp\left[-2\gamma\int_{0}^{t}ds\,\varphi(s)\right]\right\}$, whereas  $b(t)=8K\int_{0}^{t}ds\,e^{-2\gamma(t-s)}\varphi(s)$ when $\gamma(t)=\gamma$.

Armed with the probability distribution $Q(L_{1},L_{2},t)$ one can now write down the full probability $P(x,L_{1},L_{2},t)$ because $W(x,t|L_{1},L_{2})$ can be obtained analytically with the method of images \cite{chandrasekhar1943,montrollwest1987} as
\begin{equation}
W_{x_{0}}(x,t)=\left[H(x-L_{1})-H(x-L_{2})\right]\sum_{n=-\infty}^{+\infty}\frac{e^{-\frac{\left[x+2n(L_{2}-L_{1})-x_{0}\right]^{2}}{w(t)}}+e^{-\frac{\left[x-2L_{1}+2n(L_{2}-L_{1})+x_{0}\right]^{2}}{w(t)}}}{\sqrt{\pi w(t)}},
\label{eqwalker}
\end{equation}
where $x_{0}$ is the initial walker position and $w(t)=4Dt$ with $D$ being the diffusion coefficient of the walker. The Heaviside functions in Eq. (\ref{eqwalker}) show explicitly that outside the region $L_{1}\leq x\leq L_{2}$ the walker probability distribution is identically zero. For ease of notation we have omitted here and in the rest of the paper $L_{1},L_{2}$ in the definition of $W$. The quantity directly comparable to the simulation output, which is also easily accessible from field data, is the marginal probability distribution $M(x,t)$ of the walker position $x$
\begin{equation}
M(x,t)=\int_{0}^{+\infty}d\lambda\,\mathcal{Z}(\lambda,t)\int_{-\infty}^{+\infty}d\mathcal{L}\,\,\frac{e^{-\frac{\mathcal{L}^{2}}{c(t)}}}{\sqrt{\pi c(t)}}W_{x_{0}}(x,t),
\label{marginal}
\end{equation}
wherein
\begin{equation}
\mathcal{Z}(\lambda,t)=\frac{e^{-\frac{\lambda^{2}+L^{2}}{b (t)}}}{\sqrt{\pi b(t)}}2\cosh\left(\frac{2\lambda L}{b(t)}\right).
\label{Zfunct}
\end{equation}

We study in particular the dynamics of Eq. (\ref{marginal}) for a time dependence in the diffusion constant of the form $\varphi(t)=\alpha (t/\zeta)^{\alpha-1}$ with $\zeta$ being a characteristic time. This choice allows the tuning of the anomalous diffusion of the boundaries in terms of only one parameter, the exponent $\alpha$. From the expression derived after performing the integration over $\mathcal{L}$ (see Appendix B for details) one notices that by
using the dimensionless parameters $D'=D/(L^{2}\gamma)$, $K'=K/(L^{2}\gamma)$, $\beta=\gamma\zeta$ and $\tau=\gamma t$, the time-dependence in Eq. (\ref{marginal}) is lumped into the functions $c(t)/L^{2}$, $b(t)/L^{2}$ and $w(t)/L^{2}$. $D'$ and $K'$ represent the average area covered in a time $\gamma^{-1}$ by, respectively, the walker and the boundaries relative to the square of the average boundary separation $L$. $\beta$, being proportional to $\gamma$, represents the dimensionless rate at which the boundary separation returns to its average value. For subdiffusive processes, which we focus on here, we have $0<\alpha<1$, with $\alpha=1$ being the limiting diffusive case. We thus have $b(t)/L^{2}=4K'\left[1-\exp(-2\beta^{1-\alpha}\tau^{\alpha})\right]$ when $\gamma(t)=\gamma\varphi(t)$ or $b(t)/L^{2}=8K'\beta^{1-\alpha}\int_{0}^{\tau}dp\,\exp[-2(\tau-p)]\alpha p^{\alpha-1}$ when $\gamma(t)=\gamma$, whereas $c(t)/L^{2}=2K'\beta^{1-\alpha}\tau^{\alpha}$ and for the walker we have $w(t)/L^{2}=4D'\tau$.
\begin{figure}[htb]
\includegraphics[width=\columnwidth]{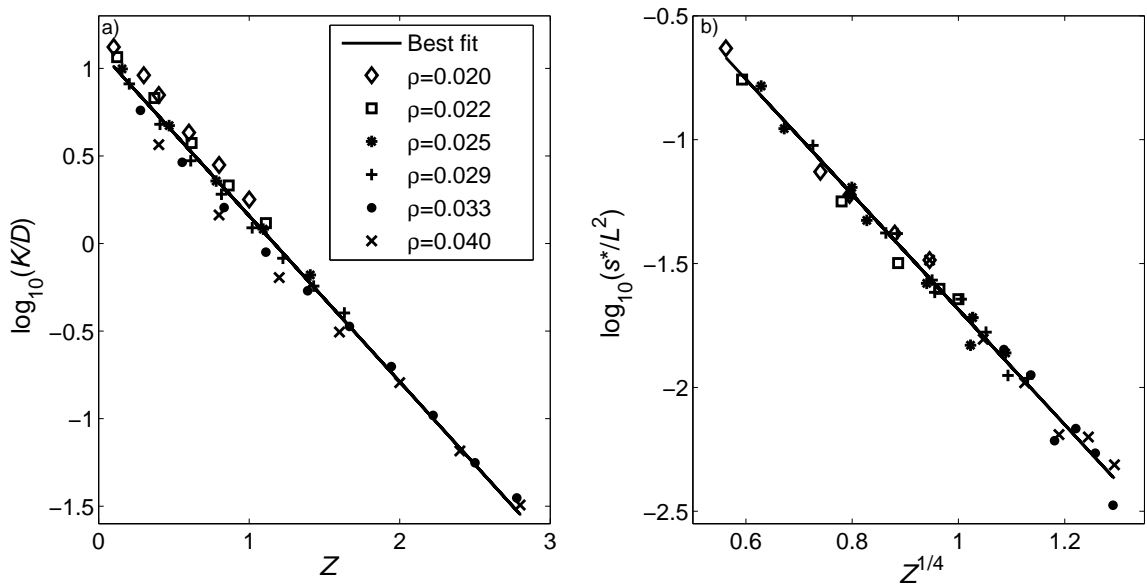}
\caption{Dependence of the parameters in the reduced model of Sec. \ref{sec2} in terms of the output from the stochastic simulations of the many-body problem. Panel (a) is a log-linear plot of the dimensionless $K/D$ versus the dimensionless quantity $Z=T_{AS}'\rho'^{2}$ where $\rho'$ is the population density multiplied by the lattice spacing and $T_{AS}'$ is the dimensionless quantity obtained by multiplying $T_{AS}$ with the random walk transfer rate $F$ between nearest neighbour sites. Panel (b) is a log-linear plot of the dimensionless quantity $s^{*}/L^{2}$ versus  $Z^{1/4}$. The solid lines are least square fits of all the data points for respectively $\mbox{Log}(K/D)= 1.11 - 0.95\times Z$ in panel (a) and $\mbox{Log}(\bar{s}/L^{2}) = 0.64 -2.32\times Z^{1/4}$ in panel (b). Each point in (a) is found by averaging over $10^5$ stochastic realisations of the simulation for sufficiently many time-steps to reach the boundary MSD's asymptotic regime. The time to reach this asymptotic regime increases exponentially with both $\rho'$ and $T_{AS}'$.  For lower values of $\rho'$, a high $T_{AS}'$ is required for the boundary MSD to reach the asymptotic regime before the limits of the box-size are reached.  Therefore using either very high or very low $\rho'$ is not practically possible, allowing us only the limited set of values in this plot.  For (b), a single simulation of $10^8$ time-steps gave sufficient data for finding each data-point.}
\label{lambdaktasrho}
\end{figure}

\section{Stochastic simulations of the many-body problem}
\label{sec3}
We compare stochastic simulations of two territorial random walkers on a ring lattice with the reduced model of Sec. \ref{sec2} in the subordinate case, i.e. when $\gamma(t)=\gamma\varphi(t)=\frac{1}{2}\gamma t^{-\frac{1}{2}}$. With this choice of $\gamma(t)$, the MSD of the boundary separation in the simulations is qualitatively similar to the reduced model, in that $\langle(L-\lambda)^2\rangle$ is monotonic, increasing initially and saturating at long times. In order to make this comparison we need to determine the dependence of $K$ and $\gamma$ on the active scent time $T_{AS}$ and the walker population density $\rho$, which in this case is simply twice the inverse of the number of lattice sites in the ring. In Fig. \ref{lambdaktasrho} we show these dependencies for a range of values of active scent time and population density by plotting the dimensionless parameter $K/D$ and the asymptotic value $s^{*}=s(t\rightarrow +\infty)/L^{2}$ of the normalized MSD of the boundary separation distance $s(t)$ (see Appendix A for its analytic expression). These parameters depend on the quantity $Z=T_{AS}'\rho'^2$ where $T_{AS}'$ and $\rho'$ are dimensionless versions of $T_{AS}$ and $\rho$ respectively (see caption of Fig. \ref{lambdaktasrho}). Qualitatively one would expect $K$ to decrease as function of $Z$ as the latter represents the time for the scent mark to disappear, $T_{AS}$, divided by the mean first passage time to move between the left and right boundaries, which is proportional to a first approximation to the length squared of the domain itself \cite{condaminetal2007}. The rate of movement of the boundaries is obviously larger the longer it takes the walker to refresh its own scent marks as well as the shorter the scent remains active. The fitting lines obtained in Fig. \ref{lambdaktasrho} provide the means for selecting the appropriate parameter values $K'$, $D'$ and $\beta$, which are present in the expression for the marginal probability distribution of the walker in Eq. (\ref{marginalw}).
\begin{figure}[htb]
\includegraphics[width=\columnwidth]{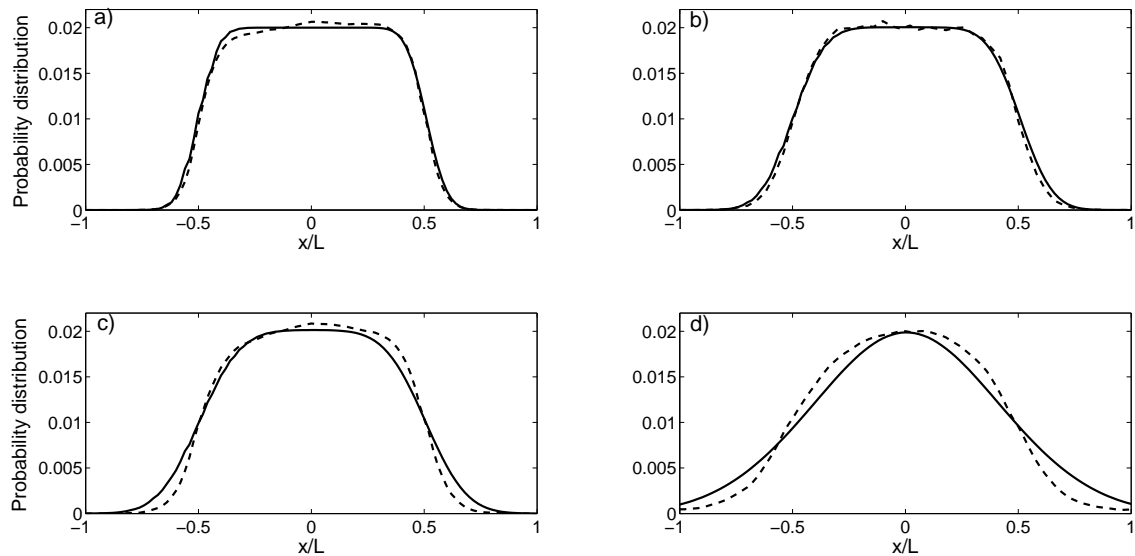}
\caption{Comparison of the walker probability distribution from the simulated model (dashed lines) with the reduced model (solid lines) at time $tF$=10,500 for four values of Z: a) Z=1.8, b) Z=1.4, c) Z=1.2 and d) Z=0.6 To plot the probability distribution of the reduced model, $M(x,t)$ in Eq. \ref{marginalw} is multiplied by the lattice constant. In all panels the density has been kept fixed at $\rho'=0.02$, whereas $T_{AS}'=9000, 7000, 6000, 3000$ for the panel a), b), c) and d), respectively. Running the simulations requires a choice of initial scent spatial profile, i.e. the time value associated with each lattice site that defines how long before the scent stops being active. We have selected a biologically relevant scent profile with minimum values at the scent boundaries and a maximum in the middle of the territory. The initial location of the two walkers is consequently the middle of their respective territories (see ref. \cite{giuggiolietal2011} for further details).}
\label{sim_vs_analytic}
\end{figure}

For a choice of $T_{AS}'$ and $\rho'$ in the simulations, one determines $K/(\gamma L^{2})$ with $L=\rho^{-1}$, and $K/D$ of the reduced model. The walker diffusion constant is simply $a^2F$ where $a$ is the lattice spacing and $F$ the random walk transfer rate between nearest neighbour sites. $F$ and $a$ do not need to be measured since they are input of the model. From Fig. \ref{lambdaktasrho}a one can now determine $K$ and subsequently $\gamma$ from Fig. \ref{lambdaktasrho}b. In Eq. (\ref{fptdc}) for convenience $K$ has units of a diffusion constant and $\varphi(t)$ is a dimensionless quantity since time is rescaled with the time constant $\zeta$. To measure $K$ from the MSD of the boundary in the simulation, it is necessary to choose a value for $\zeta$. Figure \ref{lambdaktasrho}a was produced by picking $\zeta=1/F$ so this value of $\zeta$ must be used in any comparisons between the simulation output and the reduced model.  However, this choice is arbitrary since if we were to choose $\zeta=C/F$ for some $C>0$ instead, each value of $K$ measured from the simulation output would be reduced by a factor of $C^{1-\alpha}$, i.e. points in Fig. \ref{lambdaktasrho}a would be shifted down by $\mbox{Log}(C^{1-\alpha}D)$. Then, if the new values of $\zeta$ and $K$ were placed back into Eq. (\ref{fptdc}), the changes would cancel one another out.

In Fig. \ref{sim_vs_analytic} we show the result of this exercise for four sets of $T_{AS}'-\rho'$ parameter choice by superimposing the simulated probability distribution of either walker and the marginal probability distribution of the reduced model, whose parameters have been selected from the empirically derived long-time relations between $K/D$ and $Z$ and $s^{*}/L^{2}$ and $Z$. By looking sequentially at the panels from (a) to (d) in Fig. \ref{sim_vs_analytic} it is evident that the dissimilarity between the theoretical curves and their respective simulated curves increases as $Z$ gets smaller. This is due to the assumptions in our reduced model. The adiabatic approximation relies upon the walker probability distribution to be close to the case when boundaries are immobile, a situation corresponding to $T_{AS}=\infty$. A reduction of the value of $Z$ corresponds to a decrease in the active scent time and/or a decrease of the walker population density. For a given density $\rho$ of walkers, a large reduction in $T_{AS}$ causes the region where foreign scent is present to move too quickly compared to the diffusion time of the walker. Scent deposited at a boundary site is more likely to become inactive before the animal has had a chance to re-visit that site, causing the boundary to move with higher probability. Similarly, for a given value of $T_{AS}$, a big reduction in $\rho$ makes the walkers move within a much larger area. The time it takes for the walker to roam between the territorial boundaries becomes too large to assume that $L_{1}$ and $L_{2}$ move little compared to the walker. In other words, as $Z$ diminishes the conditions for the assumptions in the adiabatic approximation breaks down.

\section{Non-monotonic dependence of the mean square displacement}
\label{negativediff}
An important quantity that characterizes the walker anomalous movement is the time-dependence of the MSD averaged over all boundaries' locations:
\begin{equation}
\langle\langle\langle (x-x_{0})^{2}\rangle\rangle\rangle(t)=\int_{-\infty}^{+\infty}dL_{2}\int_{-\infty}^{+\infty}dL_{1}\int_{L_{1}}^{L_{2}}dx\,(x-x_{0})^{2}\,Q(L_{1},L_{2},t)\,W_{x_{0}}(x,t),
\label{msd}
\end{equation}
where the symbol $\langle\langle\langle ... \rangle\rangle\rangle$ indicates that the average is over the three variables $x$, $L_{1}$ and $L_{2}$. By performing two of the integrations Eq. (\ref{msd}) can be reduced (see Appendix C for details) to the analytic formula
\begin{eqnarray}
\langle\langle\langle (x-x_{0})^{2}\rangle\rangle\rangle&&=\,\,\,x_{0}^{2}+\frac{L^{2}}{12}+\frac{b(t)}{24}+\frac{c(t)}{2}+\int_{0}^{+\infty}d\lambda\mathcal{Z
}(\lambda,t)\sum_{n=1}^{+\infty}\left\{\frac{4\lambda^{2}(-1)^{n}}{\pi^{2}(2n)^{2}}\cos\left(\frac{2n\pi x_{0}}{\lambda}\right)e^{-\frac{(2n)^{2}\pi^{2}\left[c(t)+w(t)\right]}{4\lambda^{2}}}\right. \nonumber \\
&&+\,\,\left.\frac{4(-1)^{n}}{\pi(2n-1)} \left[c(t)\cos\left(\frac{(2n-1)\pi x_{0}}{\lambda}\right)+\frac{2\lambda x_{0}}{\pi(2n-1)}\sin\left(\frac{(2n-1)\pi x_{0}}{\lambda}\right)\right]e^{-\frac{\pi^{2}(2n-1)^{2}\left[c(t)+w(t)\right]}{4\lambda^{2}}}\right\},
\label{msdeqtdc}
\end{eqnarray}
which is a generalization of the time-dependence of a Brownian walker MSD within immobile boundaries whose distance equals $L$. In the limit $K\rightarrow 0$ we have in fact that $c(t)\rightarrow 0$ and $b(t)\rightarrow 0$ which transforms $\mathcal{Z}(\lambda,t)$ into $\delta(\lambda -L)$, reducing Eq. (\ref{msdeqtdc}) to a well known expression (see e.g. ref. \cite{giuggiolietal2005}).

An inspection of Eq. (\ref{msdeqtdc}) when $x_{0}=0$ shows that by expanding the exponential up to second order in the even series and up to first order in the odd series one obtains $\langle\langle\langle x^{2}\rangle\rangle\rangle\sim w(t)$ for $t\rightarrow 0$. The short-time dependence of the MSD is thus dependent only on the walker movement when away from the boundaries. The long time behaviour on the other hand shows that $\langle\langle\langle x^{2}\rangle\rangle\rangle\sim L^{2}/12+b(t)/24+c(t)/2$. The two series in Eq. (\ref{msdeqtdc}) in fact decay to zero when $t\rightarrow +\infty$ because of the exponential terms. In other words the boundary dispersal dictates how the MSD increases at long times through the diffusion of the boundary separation $b(t)$ and the boundary centroid $c(t)$. Since $b(t)$ and $c(t)$ contains $\varphi(t)$, the choice of the exponent $\alpha$ allows us to control the time-dependence of the (normalized) asymptotic factor $a(t)=1+b(t)/(2L^{2})+6c(t)/L^{2}$ and consequently the long-time anomalous diffusion characteristics of the MSD. For the case $\gamma(t)=\gamma\varphi(t)$, $b(t)$ reaches a positive asymptotic value, as shown previously, whereas $b(t)$ approaches 0 at long times when $\gamma(t)=\gamma$. This means that in both cases the dominant effect in the walker MSD at long times is due to the boundary centroid rather than the boundary separation. In the remaining part of this section we focus on the case $\gamma(t)=\gamma$.

The non-monotonicity of the time-dependence in the $b(t)$ under certain conditions may also appear in $a(t)$. When that occurs we say that the walker undergoes the phenomenon of negative diffusion as a result of the interplay between two contributions to the boundary movement: the centroid displacement, governed by $c(t)$, and the deviation of the boundary separation distance from $L$, governed by $b(t)$. The former is monotonically increasing, whereas the latter is not, increasing initially, before reaching a peak and decaying to zero at long times. If at any point the increase of $c(t)$ is sufficiently slow compared to the decrease of $b(t)$, the boundary MSD will decrease.
\begin{figure}[htb]
\includegraphics[width=\columnwidth]{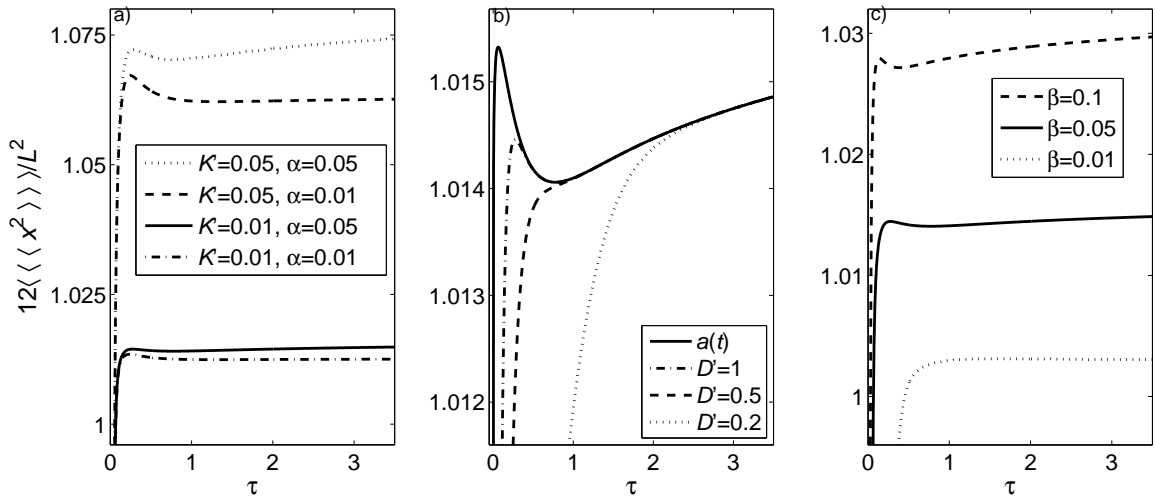}
\caption{Non-monotonicity of the walker's MSD (Eq. (\ref{msdeqtdc}) with $x_{0}=0$) for different parameter values. The subfigure (a) shows how the non-monotonic behaviour of the MSD, for two choices of $\alpha<\alpha_{0}$, tends to be washed away as $K'$ decreases, while the $D'$ and $\beta$ are kept fixed. In the subfigure (b) the MSD dependence over $D'$ is shown while keeping the other parameters fixed: $K'=0.01$, $\alpha=0.05$, $\beta=0.1$, and in subfigure (c) the MSD dependence over $\beta$ is shown with $K'=0.01$, $\alpha=0.05$, $D'=1$.}
\label{MSDplots}
\end{figure}

The parameter dependence for this non-monotonicity can be found by first studying the time dependence of the walker's MSD asymptotic expression $a(t)$, which possesses a maximum and a minimum whenever the $\alpha$-dependent function $v_{\alpha}(\tau)>2$ with $v_{\alpha}(\tau)=\tau^{1-\alpha}\int_{0}^{\tau}ds\,\exp[-2(\tau-s)] s^{\alpha-1}$. A numerical study of $v_{\alpha}(\tau)$ shows that this occurs when $\alpha<\alpha_{0}$ with $\alpha_{0}\approx 0.105$ and irrespective of any other parameters. In Fig. \ref{MSDplots} we show a few examples with $\alpha<\alpha_{0}$ where the MSD possesses a dip in time for different values of the parameters $K'$, $D'$, $\beta$ and $\alpha$. Fig. \ref{MSDplots}a shows that, for fixed $D'$, $\beta$ and $\alpha$, decreasing $K'$ tends to smooth out the appearance of the dip, eventually making it disappear with further decrease of $K'$ (not shown in the figure). On the other hand, maintaining $K'$, $\beta$ and $\alpha$ fixed, it is the increase of $D'$ that eventually makes the dip disappear, as one can see by looking sequentially at the various curve in Fig. \ref{MSDplots}b. A qualitative similar dependence is depicted in Fig. \ref{MSDplots}c as $\beta$ decreases while keeping the other parameters fixed.

To explain why varying $K'$, $D'$ and $\beta$ as shown in Fig. \ref{MSDplots} make the transient negative diffusivity present in the expression $a(t)$ disappear, we rewrite Eq. (\ref{msdeqtdc}). We do so by performing the $\lambda$ integration (see Appendix D for details). In the resulting expression (\ref{msdeqtdc2}), besides the asymptotic term $a(t)$, the time dependence is lumped into powers of order $f(t)^{k/2}$ with $k>1$, but more importantly terms like $\exp[-L^{2}/b(t)]$, $\exp[-2\pi\sqrt{f(t)}]$ and $\exp[-\pi\sqrt{f(t)}]$ containing the dimensionless quantity $f(t)=[c(t)+w(t)]/b(t)=\left(\tau^{\alpha}+2D'\tau\beta^{\alpha-1}/K'\right)/\left\{4\int_{0}^{\tau}dp\,\exp[-2(\tau-p)]\alpha p^{\alpha-1}\right\}$. Since $f(t)$ is linearly proportional to $D'$ and inversely proportional to $K'\beta^{1-\alpha}$, whereas $L^{2}/b(t)$ is inversely proportional to $K'\beta^{1-\alpha}$, it is possible to understand qualitatively the direct and inverse proportionality on $K'$ and $D'$ that can be evinced by looking sequentially at the various curves in Fig. \ref{MSDplots}. As $D'$ increases, the value of $f(t)$ increases, and the MSD more rapidly becomes very close to its asymptotic expression $a(t)$, as clearly shown by the trend in the plots depicted in Fig. \ref{MSDplots}a. The opposite trend is observed when $K'\beta^{1-\alpha}$ increases due to its inverse proportionality in the MSD expression.
\begin{figure}[htb]
\includegraphics[width=\columnwidth]{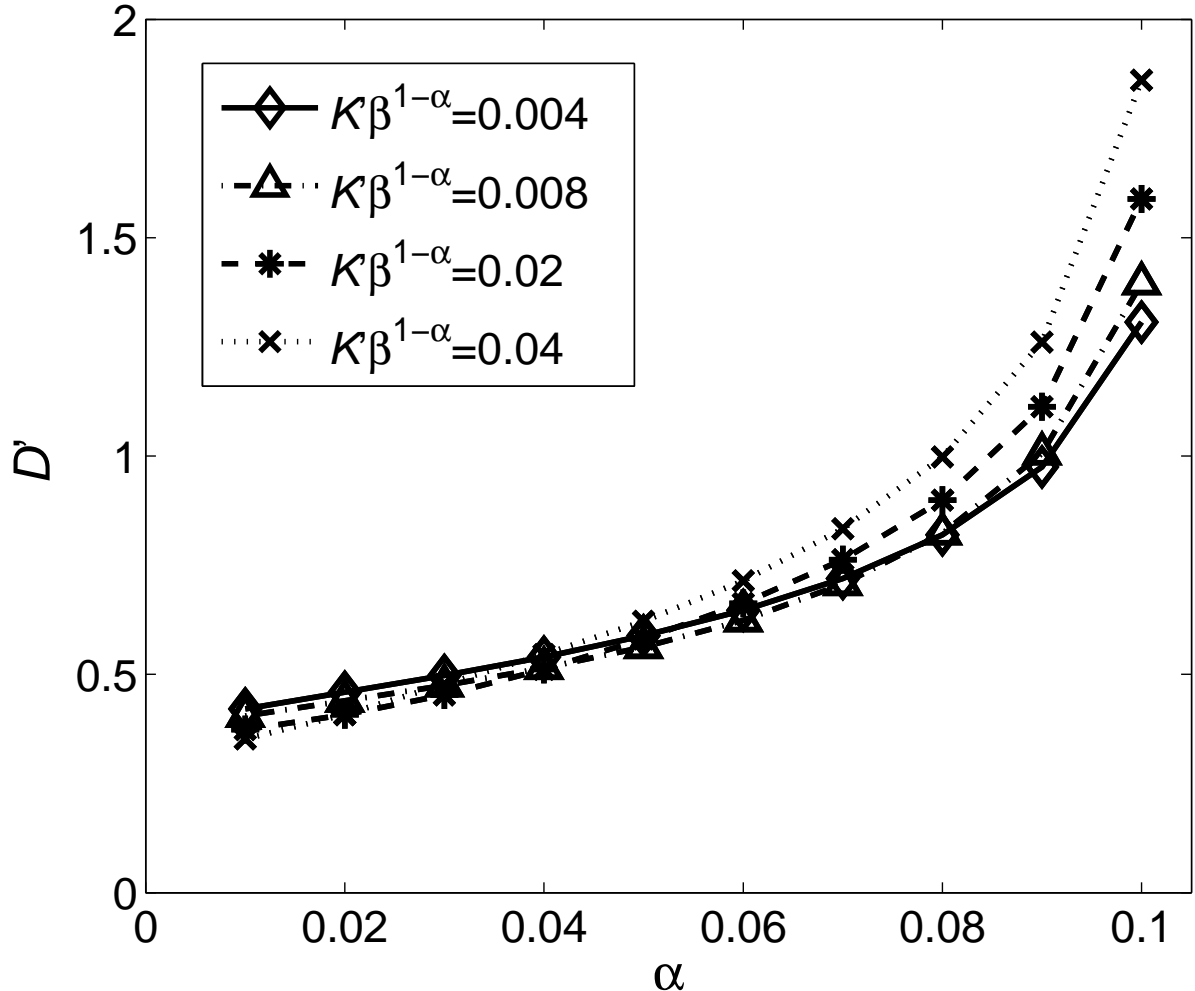}
\caption{Lines in parameter space $D'-\alpha$, for different values of $K'\beta^{1-\alpha}$, delineating the regions below which the walker's MSD is non-monotonic.}
\label{maxpos}
\end{figure}

Having identified the parameter combination for which the MSD time-dependence is non-monotonic, we now try to interpret it, and in particular we try to explain the reasons for the sudden transition at $\alpha<\alpha_{0}$. For that, it is sufficient to consider the asymptotic expression $a(t)L^{2}/12$ of the MSD and determine from it the value of the (time-dependent) diffusion constant. With simple algebra one can rewrite $d/dt[a(t)L^{2}/12]$ as an effective diffusion coefficient $D_{eff}(t)=K\phi(t)\{1+\Gamma[\alpha]E_{1,\alpha}(-2\gamma t)/3\}$ where $E_{\sigma,\delta}(z)=\sum_{n=0}^{+\infty}z^{n}/\Gamma(\sigma n+\delta)$ is the 2-parameter Mittag-Leffler function. $K\phi(t)$, the first term in $D_{eff}(t)$, is the time-dependent diffusion coefficient of either boundary, whereas the second term in $D_{eff}(t)$ is the one for the boundary separation. Interestingly, the function $\Gamma[\alpha]E_{1,\alpha}(-2\gamma t)/3$ starts at 1/3 at $t=0$ and remains positive for any values of time only if $\alpha=1$, i.e. when the Mittag-Leffler reduces to an exponential. For $\alpha<1$ on the other hand, $\Gamma[\alpha]E_{1,\alpha}(-2\gamma t)/3$ becomes negative, reaches a minimum and eventually approaches the value zero from below as $\sim\Gamma(\alpha)[6\Gamma(\alpha-1)\gamma t]^{-1}$.

The negative diffusion is due to the qualitative difference in time scales of two processes: the random movement of the boundary separation variable $\lambda$ and its rate of return to the value $L$. The former is sublinear, except when $\alpha=1$, whereas the latter is always linear. At short times the Gaussian nature of the process makes the effective diffusion constant of the boundary separation positive. Although at time $t=0$ we have set the random variable $\lambda$ to be $\delta$-localized, i.e. zero everywhere except at $\lambda=L$, at any time $t>0$, $\lambda$ can assume values much larger than $L$ due to the Gaussian shape of $\mathcal{Q}_{2}(\lambda,t)$. Irrespective of the shape of the constraining potential, the diffusion constant for the boundary separation is initially positive making the MSD increase sub-linearly (when $\alpha<1$). The spring force acts against this initial transient increase in the MSD by pushing the value $\lambda$ back to $L$. The drift towards $L$ occurs on such a fast time scale that it counteracts the boundary separation dispersion, making the MSD decrease its value after a certain amount of time. At that moment the system can be described by an effective diffusion constant with a negative sign.

Although the above description is valid for any $\alpha<1$, the transition is observed only at $\alpha=\alpha_{0}$ rather than $\alpha=1$. This is because we have only described the dynamics of the boundary separation, neglecting the effects of the boundary centroid. The decreasing trend observed in the full asymptotic expression $a(t)$ eventually ends once the reduction in the MSD of the boundary separation is overcome by the monotonic increase in the MSD of the boundary centroid. The effective diffusion constant $D_{eff}(t)$ is in fact the sum of an always positive contribution due to the boundary centroid, and a contribution due to the boundary separation, which is initially positive and negative afterwards. When one considers both contributions, it turns out that only when the boundary subdiffusion is particularly pronounced, i.e. when $\alpha<\alpha_{0}\approx 0.105$, does the negative diffusion due to the spring force fail to compensate for the positive diffusion of the boundary centroid, making the MSD non-monotonic.

\section{Conclusions}
\label{sec5}

We have analyzed the movement of a 1D Brownian walker within anomalously diffusing boundaries, focusing on the case in which each boundary moves subdiffusively at all times. We have shown how the analytic dependence of the walker MSD depends on the movement statistics of the boundary separation and boundary centroid by using a Fokker-Planck formalism with a time-dependent diffusion coefficient. The study has been motivated by the need to derive an approximate model of the many-body phenomenon of territorial emergence consisting of walkers interacting through a renormalized exclusion process for which individuals are excluded from an area recently visited by other walkers \cite{giuggiolietal2011}.

Despite the non-Markovian nature of territorial dynamics, we have exploited the time scale disparity between the territory movement, which is subdiffusive, and the walker movement, which is diffusive, to construct a reduced one-body model that captures the salient features of the many-body problem. Through an adiabatic approximation we have shown how the analytic model of the single walker moving within subdiffusing boundaries is able to represent the complex dynamics present in animal territoriality. A comparison between the probability distribution of the analytic model and the stochastic simulations of two territorial random walkers on a ring has confirmed the validity of the adiabatic approximation. 

Explicit expressions of the walker MSD have been computed, which are direct generalizations of the MSD of random walker within fixed boundaries. An interesting outcome of our formalism is the appearance, under certain conditions, of a non-monotonic behaviour of the Brownian walker MSD at intermediate times, that is to say, the system undergoes the phenomenon of transient negative diffusion (seen in other context e.g. \cite{karpov1995}).

Although a direct application of our equations is the displacement of an animal within its own moving territorial boundaries \cite{giuggiolietal2011}, any situation where the movement of two or more objects are separated by an enclosing interface or boundaries (see e.g. \cite{nelsonetalbook1989}) could benefit from our study. We have focused on the case of a Brownian walker within subdiffusing boundaries, but our formalism could also be employed in other applications in which the boundary movement is modelled as a subordinate anomalous diffusion process, by choosing more general dependence of the boundary time-dependent diffusion constant $\varphi(t)$. Similarly, one could take into account anomalously diffusing walkers by having $w(t)=4D\int_{0}^{t}ds\,\chi(s)$ where $D\chi(t)$ is a time-dependent diffusion constant.

Applications of these kinds may occur at much smaller spatial scales compared to those typical of animal territorial dynamics. At the sub-cellular level for example, the positioning of certain cell components is crucial to its functioning and relies upon the continuous exploration through a variety of motor proteins of the slowly changing intracellular space \cite{tolic-norrelykke2010}. The moving boundaries in this context would represent the translation and deformation of the cell membrane.
Examples at the cellular level can be found in developmental studies on certain macrophages that move via contact-inhibition interaction, a form of repulsion mechanism upon contact. This mechanism allows the macrophages to partition the space they continuously monitor for the presence of infectious agents \cite{strameretal2010}. This space partitioning creates effective boundaries between macrophages that continuously change in time, which could also be modeled though the help of the formalism developed here.

We have chosen to represent anomalous diffusion via a subordinate Gaussian process, but other representations are possible, one such being a time non-local formalism given by the Generalized Master Equation (GME) \cite{kenkreGME1977,kenkre2003} or its equivalent, the Fractional Diffusion Equation (FDE) \cite{metzlerklafterreview2000}  (see e.g. ref. \cite{giuggiolietal2009} for the equivalence between GME and FDE). A comparative investigation between the time local formalism used here, i.e. $\partial Q(\vec{z},t)/\partial t=K\varphi (t)\nabla^{2} Q(\vec{z},t)$, and the time non-local GME formalism, i.e. $\partial Q(\vec{z},t)/\partial t=K\int_{0}^{t} ds\,\phi(t-s)\nabla^{2}Q(\vec{z},s)$, where $\phi(t)$ is the so-called memory function, has been carried out in ref. \cite{kenkresevilla2007}. That comparative analysis, however, did not consider the presence of a constraining potential. We plan to extend that work by performing a comparative analysis between a time local and a time non-local representation of the boundary fluctuations in presence of a constraining force, as well as analyzing the influence that the potential shape has on the results presented here.

\acknowledgments
We thank Paul Martin and Brian Stramer for useful discussions. This work was partially supported by the EPSRC grant number EP/E501214/1 (L.G. and J.R.P) and the Dulverton Trust (S.H.). We acknowledge the Advanced Computing Research Centre, University of Bristol - http://www.bris.ac.uk/acrc/ - for making their computing resources available to us.

\section*{Appendix A}
\label{appA}

The variable transformation to the boundary separation $\lambda$ and boundary centroid $\mathcal{L}$ described in Sec. \ref{sec2} allows one to rewrite Eq. (\ref{fptdc}) as
\begin{equation}
\frac{\partial \mathcal{Q}(\mathcal{L},\lambda,t)}{\partial t}=K\varphi(t)\left(\frac{1}{2}\frac{\partial^{2}}{\partial \mathcal{L}^2}+2\frac{\partial^{2}}{\partial \lambda^2}\right)\mathcal{Q}(\mathcal{L},\lambda,t)+\gamma(t)\frac{\partial}{\partial \lambda}\left[(\lambda-L)\mathcal{Q}(\mathcal{L},\lambda,t)\right],
\label{eqtransf}
\end{equation}
and the corresponding initial conditions as $\mathcal{Q}(\mathcal{L},\lambda,0)=\delta(\mathcal{L}-\mathcal{L}_{0})\delta(\lambda-\lambda_{0})$ with $\mathcal{L}_{0}=(L_{1,0}+L_{2,0})/2$ and $\lambda_{0}=L_{2,0}-L_{1,0}$.

Eq. (\ref{eqtransf}) can now be decoupled as $Q(L_{1},L_{2},t)=\mathcal{Q}_{1}(\mathcal{L},t)\mathcal{Q}_{2}(\lambda,t)$, into a differential equation for the boundary separation $\lambda$
\begin{equation}
\frac{\partial \mathcal{Q}_{1}(\mathcal{L},t)}{\partial t}=\frac{K}{2}\varphi(t)\frac{\partial^{2} \mathcal{Q}_{1}(\mathcal{L},t)}{\partial \mathcal{L}^2},
\label{eqsum}
\end{equation}
and a differential equation for the boundary centroid location $\mathcal{L}$
\begin{equation}
\frac{\partial \mathcal{Q}_{2}(\lambda,t)}{\partial t}=2K\varphi(t)\frac{\partial^{2} \mathcal{Q}_{2}(\lambda,t)}{\partial \lambda^2}+\gamma(t)\frac{\partial}{\partial \lambda}\left[(\lambda-L)\mathcal{Q}_{2}(\lambda,t)\right].
\label{eqdiff}
\end{equation}
Eq. (\ref{eqsum}) and (\ref{eqdiff}) are solved together with the reflection condition at $\lambda=0$, i.e.
\begin{equation}
\frac{\partial}{\partial \lambda}\mathcal{Q}_{2}(\lambda,t)|_{\lambda=0}=0.
\label{eqbounddiff}
\end{equation}

The solution of Eq. (\ref{eqsum}) trivially gives a Gaussian whose width increases in time proportionally to $c(t)$, whereas the solution of Eq. (\ref{eqdiff}) is more involved as it needs to be solved with the method of characteristics \cite{moonspencerbook1969}. The general solution of Eq. (\ref{eqdiff}) is given by
\begin{equation}
\mathcal{Q}_{2}(\lambda,t)=\frac{e^{-\frac{(\lambda-\bar{\lambda}(t))^{2}}{b(t)}}}{\sqrt{\pi b(t)}}
\end{equation}
where
\begin{eqnarray}
\bar{\lambda}(t)&=&L+e^{-G(t)}(\lambda_{0}-L), \nonumber \\
b(t)&=&8K\int_0^t ds \varphi(s)e^{-2\left(G(t)-G(s)\right)} \\
\label{solmsigmaeq}
\end{eqnarray}
and $G(t)$ is defined so that $G'(t)=\gamma(t)$.  Once again with the help of the method of images \cite{chandrasekhar1943} one takes into account the reflective boundary condition at $\lambda=0$ and obtains the general solution of Eq. (\ref{eqdiff}) and (\ref{eqbounddiff}) as
\begin{equation}
\mathcal{Q}_{2}(\lambda,t)=H(\lambda)\frac{e^{-\frac{(\lambda-\bar{\lambda}(t))^{2}}{b(t)}}+e^{-\frac{(\lambda+\bar\lambda(t))^{2}}{b(t)}}}{\sqrt{\pi b(t)}}.
\label{eqlambda}
\end{equation}
Eq. (\ref{eqlambda}) represents the probability distribution of the positive distance $\lambda$ and it is given by the sum of two Gaussian curves with the same width but centered around two different values: $L$ for the first one and $-L$ for the second one (for the choice of initial conditions we have that $\bar\lambda(t)=L$).
If $\gamma(t)=\gamma \varphi(t)$ the width of each of these Gaussian curves starts at 0 and then increases monotonically to the value $4K/\gamma$ if $\gamma(t)=\gamma \varphi(t)$. On the other hand when $\gamma(t)=\gamma$, these curve widths have an initial increase to a maximum followed by a decrease back to 0. Given that the second Gaussian is centered at some negative value of $L$, for values of $L$ that are not too small one can think of the dynamics of $\mathcal{Q}_{2}(\lambda,t)$ as given essentially by the spreading of the Gaussian centered around the positive constant $L$. The deviations from this qualitative description can be determined exactly by computing the MSD $s(t)=\langle (\lambda-L)^{2}\rangle$ of $\lambda$ from $L$. A simple integration gives
\begin{equation}
s(t)=\frac{b(t)}{2}\left\{1-\frac{4}{\sqrt{\pi}}\frac{L}{\sqrt{b(t)}}e^{-\frac{L^{2}}{b(t)}}+4\frac{L^{2}}{b(t)}\left[1-\mbox{erf}\left(\frac{L}{\sqrt{b(t)}}\right)\right]\right\},
\label{stfunc}
\end{equation}
which shows that the MSD equals $b(t)/2$, as one would expect for the case of a Gaussian with variance $b(t)/2$ and mean $L$, only when $L/\sqrt{b(t)}>>1$. When comparing with simulations, in the case $\gamma(t)=\gamma\varphi(t)$, the pertinent value is the limit $s(t\rightarrow +\infty)=s^{*}$.  To find this limit, simply substitute $4K/\gamma$ for $b(t)$ everywhere in Eq. (\ref{stfunc}).

\section*{Appendix B}
\label{appB}
The marginal probability distribution of the walker can be computed via the double integral expression
\begin{equation}
M(x,t)=\int_{0}^{+\infty}d\lambda\,\mathcal{Z}(\lambda,t)\int_{-\infty}^{+\infty}d\mathcal{L}\,\,\frac{e^{-\frac{\mathcal{L}^{2}}{c(t)}}}{\sqrt{\pi c(t)}}g_{x_{0}}(x,t)\left[H(x-\mathcal{L}+\lambda/2)-H(x-\mathcal{L}-\lambda/2)\right],
\label{marginalw0}
\end{equation}
where $g_{x_{0}}(x,t)$ represents the series terms in Eq. (\ref{eqwalker}). By making use of the Poisson summation formula \cite{montrollwest1987} for the first series in $g_{x_{0}}(x,t)$ one can transform Eq. (\ref{marginalw0}) into
\begin{equation}
M(x,t)=\int_{0}^{+\infty}d\lambda\,\mathcal{Z}(\lambda,t)\int_{x-\lambda/2}^{x+\lambda/2}\,\,\frac{e^{-\frac{\mathcal{L}^{2}}{c(t)}}}{\sqrt{\pi c(t)}}\left\{\frac{1}{2\lambda}+\frac{1}{\lambda}\sum_{n=1}^{+\infty}\cos\left(n\pi\frac{x-x_{0}}{\lambda}\right)e^{-\frac{\pi^{2}n^{2}w(t)}{4\lambda^{2}}}+\sum_{n=-\infty}^{+\infty}\frac{e^{-\frac{\left[x-2\mathcal{L}+(2n+1)\lambda+x_{0}\right]^{2}}{w(t)}}}{\sqrt{\pi w(t)}}\right\}.
\end{equation}
Integration over $\mathcal{L}$ then gives
\begin{eqnarray}
M(x,t)&=&\frac{1}{2}\int_{0}^{+\infty}d\lambda\,\mathcal{Z}(\lambda,t)\left\{\left[\mbox{erf}\left(\frac{x+\lambda/2}{\sqrt{c(t)}}\right)-\mbox{erf}\left(\frac{x-\lambda/2}{\sqrt{c(t)}}\right)\right]\left[\frac{1}{2\lambda}+\frac{1}{\lambda}\sum_{n=1}^{+\infty}\cos\left(n\pi\frac{x-x_{0}}{\lambda}\right)e^{-\frac{\pi^{2}n^{2}w(t)}{4\lambda^{2}}}\right]+\right.\nonumber \\
&&\left.\sum_{n=-\infty}^{+\infty}\frac{e^{-\frac{[x+x_{0}+(2n+1)\lambda]^{2}}{4c(t)+w(t)}}}{\sqrt{\pi[4c(t)+w(t)]}}\left(\mbox{erf}\left[\frac{h(x,x_{0},t)+\lambda\,q^{-}_{n}(t)}{\sqrt{4c(t)+w(t)}}\right]-\mbox{erf}\left[\frac{h(x,x_{0},t)-\lambda\,\left(q^{+}_{n}(t)+4\sqrt{\frac{c(t)}{w(t)}}\right)}{\sqrt{4c(t)+w(t)}}\right]\right)\right\}, \nonumber \\
\left.\right.
\label{marginalw}
\end{eqnarray}
with
\begin{eqnarray}
h(x,x_{0},t)&=&x\left(\sqrt{\frac{w(t)}{c(t)}}+2\sqrt{\frac{c(t)}{w(t)}}\right)-2x_{0}\sqrt{\frac{c(t)}{w(t)}}, \nonumber \\
q^{\pm}_{n}(t)&=&\frac{1}{2}\sqrt{\frac{w(t)}{c(t)}}\pm4n\sqrt{\frac{c(t)}{w(t)}}.
\label{hqpm}
\end{eqnarray}

\section*{Appendix C}
\label{appC}

For a simplified analytic expression of Eq. (\ref{msd}), it is convenient to rewrite each infinite series in Eq. (\ref{eqwalker}) by making use of the Poisson summation formula \cite{montrollwest1987}, which gives
\begin{eqnarray}
g_{x_{0}}(x,t)&=&\frac{1}{\lambda}+\frac{1}{\lambda}\sum_{n=1}^{+\infty}\left[\cos\left(n\pi\frac{x-x_{0}}{\lambda}\right)+(-1)^{n}\cos\left(n\pi\frac{x+x_{0}}{\lambda}\right)\cos\left(2n\pi \frac{\mathcal{L}}{\lambda}\right)\right. \nonumber \\
&&\left.+(-1)^{n}\sin\left(n\pi\frac{x+x_{0}}{\lambda}\right)\sin\left(2n\pi \frac{\mathcal{L}}{\lambda}\right) \right] e^{-\frac{\pi^{2}n^{2}w(t)}{4\lambda^{2}}}.
\label{eqwalker2}
\end{eqnarray}

To obtain Eq. (\ref{msdeqtdc}) from (\ref{msd}) in the main text, one can perform the integration over $x$ between the values $L_{1}$ and $L_{2}$, and subsequently the $\mathcal{L}$-integration after transforming to $(\lambda,\mathcal{L})$-space. Alternatively one can exploit the relation $\langle (x-x_{0})^{2} \rangle=\left[-\frac{\partial^{2}}{\partial k^{2}}\mathcal{F}\{W_{x_{0}}(x,t)\}+2ix_{0}\frac{\partial}{\partial k}\mathcal{F}\{W_{x_{0}}(x,t)\}+x_{0}^{2}\mathcal{F}\{W_{x_{0}}(x,t)\}\right]|_{k=0}$, where $\mathcal{F}\{W_{x_{0}}(x,t)\}$ is the Fourier transform of $W_{x_{0}}(x,t)$. We write here the details of this calculation due to its potential applicability when only the Fourier expression of $W_{x_{0}}(x,t)$ is known, e.g. when the walker probability is a L\'{e}vy distribution.

Since $W_{x_{0}}(x,t)$ is the product of $g_{x_{0}}(x,t)$ and the step function $H(x-L_{1})-H(x-L_{2})$, the Fourier expression of $W_{x_{0}}(x,t)$ is the convolution of $\mathcal{F}\{W_{0}(x,t)\}$ with $\mathcal{F}\{H(x-L_{1})-H(x-L_{2})\}$ which are equal, respectively, to
\begin{eqnarray}
\mathcal{F}\left\{g_{x_{0}}(x,t)\right\}(k)&=&\frac{2\pi}{\lambda}\delta(k)+\frac{\pi}{\lambda}\sum_{n=1}^{+\infty}\left\{e^{i\frac{n\pi x_{0}}{\lambda}}\delta\left( k-\frac{n\pi}{\lambda}\right)+e^{-i\frac{n\pi x_{0}}{\lambda}}\delta\left( k+\frac{n\pi}{\lambda}\right)\right. \nonumber \\
&&\left.+(-1)^{n}\left[e^{i\frac{n\pi (2\mathcal{L}-x_{0})}{\lambda}}\delta\left( k-\frac{n\pi}{\lambda}\right)+e^{-i\frac{n\pi (2\mathcal{L}-x_{0})}{\lambda}}\delta\left( k+\frac{n\pi}{\lambda}\right)\right]\right\}e^{-\frac{\pi^{2}n^{2}w(t)}{4\lambda^{2}}}, 
\label{eqwalkerFour}
\end{eqnarray}
and
\begin{equation}
\mathcal{F}\left\{H(x-L_{1})-H(x-L_{2})\right\}(k)=\frac{2e^{ik\mathcal{L}}\sin\left(\frac{k\lambda}{2}\right)}{k}.
\label{fourstep}
\end{equation}
By exploiting the presence of the various Dirac delta function in (\ref{eqwalkerFour}), one can compute the convolution in Fourier domain between Eq. (\ref{eqwalkerFour}) and (\ref{fourstep}) and obtain
\begin{eqnarray}
&&\mathcal{F}\left\{W_{x_{0}}(x,t)\right\}(k)=\frac{2e^{ik\mathcal{L}}\sin\left(\frac{k\lambda}{2}\right)}{k\lambda}+\frac{1}{\lambda}\sum_{n=1}^{+\infty}\left[\frac{e^{i\frac{n\pi x_{0}}{\lambda}}e^{i\left(k-\frac{n\pi}{\lambda}\right)\mathcal{L}}\sin\left[\left(k-\frac{n\pi}{\lambda}\right)\frac{\lambda}{2}\right]}{k-\frac{n\pi}{\lambda}}+\right.\nonumber \\
&&\left. +\frac{e^{-i\frac{n\pi x_{0}}{\lambda}}e^{i\left(k+\frac{n\pi}{\lambda}\right)\mathcal{L}}\sin\left[\left(k+\frac{n\pi}{\lambda}\right)\frac{\lambda}{2}\right]}{k+\frac{n\pi}{\lambda}}+(-1)^{n}\left(\frac{e^{i\frac{n\pi (2\mathcal{L}-x_{0})}{\lambda}}e^{i\left(k-\frac{n\pi}{\lambda}\right)\mathcal{L}}\sin\left[\left(k-\frac{n\pi}{\lambda}\right)\frac{\lambda}{2}\right]}{k-\frac{n\pi}{\lambda}}\right.\right.\nonumber \\
&&\left. \left.+\frac{e^{-i\frac{n\pi (2\mathcal{L}-x_{0})}{\lambda}}e^{i\left(k+\frac{n\pi}{\lambda}\right)\mathcal{L}}\sin\left[\left(k+\frac{n\pi}{\lambda}\right)\frac{\lambda}{2}\right]}{k+\frac{n\pi}{\lambda}}\right)\right]e^{-\frac{\pi^{2}n^{2}w(t)}{4\lambda^{2}}}.
\label{eqwalkerFour2}
\end{eqnarray}
From the Fourier expression $\mathcal{F}\{W_{x_{0}}(x,t)\}$ in (\ref{eqwalkerFour2}) it is now straightforward to determine that
\begin{eqnarray}
\langle (x-x_{0})^{2} \rangle&=&(\mathcal{L}-x_{0})^{2}+\frac{\lambda^{2}}{12}+\sum_{n=1}^{+\infty}\frac{4\lambda^{2}(-1)^{n}}{(2n)^{2}\pi^{2}}\cos\left[\frac{2n\pi(\mathcal{L}-x_{0})}{\lambda}\right]e^{-\frac{\pi^{2}(2n)^{2}w(t)}{4\lambda^{2}}} \nonumber \\ 
&&+8\lambda(\mathcal{L}-x_{0})\sum_{n=1}^{+\infty}\frac{(-1)^{n}}{(2n-1)^{2}\pi^{2}}\sin\left[\frac{(2n-1)\pi(\mathcal{L}-x_{0})}{\lambda}\right]e^{-\frac{\pi^{2}(2n-1)^{2}w(t)}{4\lambda^{2}}}
\label{simpavemsd}
\end{eqnarray}
After inserting Eq. (\ref{simpavemsd}) in Eq. (\ref{msd}), changing the $L_{1}$- and $L_{2}$-integration into a $\lambda$- and $\mathcal{L}$-integration and performing the $\mathcal{L}$-integration, one then recovers Eq. (\ref{msdeqtdc}) in the main text. 

\section*{Appendix D}
\label{appD}
When $x_{0}=0$ the MSD expression (\ref{msdeqtdc}) can be simplified further if the infinite summation and the integral can be interchanged, which occurs when the series is uniformly convergent \cite{knoppbook1990} (this occurs for all times when $\gamma(t)=\gamma\varphi(t)$, whereas it occurs for $t>0$ when $\gamma(t)=\gamma$). One first needs to notice that
\begin{equation}
\int_{0}^{+\infty}d\lambda\,\,\lambda^{2m}e^{-\beta\lambda^{2}-\frac{\alpha^{2}}{\lambda^{2}}}=\frac{\sqrt{\pi}}{2}(-1)^{m}\frac{d^{m}}{d\beta^{m}}\left(\frac{e^{-2\alpha\sqrt{\beta}}}{\sqrt{\beta}}\right).
\label{lambdaint1}
\end{equation}
When Eq. (\ref{lambdaint1}) is evaluated at $\beta=1$ one obtains
\begin{equation}
\int_{0}^{+\infty}d\lambda\,e^{-\lambda^{2}-\frac{\alpha^{2}}{\lambda^{2}}}\lambda^{2m}=\frac{\sqrt{\pi}}{2^{m+1}}\Theta_{m}(2\alpha)e^{-2\alpha},
\label{lamdaint2}
\end{equation}
where
\begin{equation}
\Theta_{m}(z)=\frac{1}{2^{m}}\sum_{k=0}^{m}\frac{(2m-k)!(2z)^{k}}{(m-k)!k!}
\end{equation}
are Bessel polynomials \cite{grosswaldbook1978}. By using the series representation of the hyperbolic cosine one can now perform the $\lambda$-integration and obtain
\begin{eqnarray}
\langle\langle\langle  x^{2}\rangle\rangle\rangle&=&\frac{L^{2}}{12}+\frac{b(t)}{24}+\frac{c(t)}{2}+e^{-\frac{L^{2}}{b (t)}}\sum_{m=0}^{+\infty}\left(\frac{4L^{2}}{b(t)}\right)^{m}\frac{1}{(2m)!}\left[\frac{b(t)}{\pi^{2}}\sum_{n=1}^{+\infty}\frac{\Theta_{m+1}\left(2\pi n\sqrt{f(t)}\right)}{2^{m+1}}\frac{(-1)^{n}}{n^{2}}e^{-2\pi n\sqrt{f(t)}}+\right. \nonumber \\
&&\left.\frac{4c(t)}{\pi} \sum_{n=1}^{+\infty}\frac{\Theta_{m}\left(\pi (2n-1)\sqrt{f(t)}\right)}{2^{m}}\frac{(-1)^{n}}{2n-1}e^{-\pi (2n-1)\sqrt{f(t)}}\right],
\label{setp1}
\end{eqnarray}
with $f(t)=[c(t)+w(t)]/b(t)$ as defined in the main text. The summation over $n$ can now be performed by rewriting the two series as, respectively,
\begin{equation}
\sum_{n=1}^{+\infty}(-1)^{n}n^{k-2}e^{-2\pi n\sqrt{f(t)}}=\sum_{n=1}^{+\infty}\frac{\left(-e^{-2\pi \sqrt{f(t)}}\right)^{n}}{n^{2-k}},
\label{step2}
\end{equation}
and
\begin{equation}
\sum_{n=1}^{+\infty}(-1)^{n}(2n-1)^{k-1}e^{-\pi (2n-1)\sqrt{f(t)}}=-2^{k-1}e^{-\pi\sqrt{f(t)}}\sum_{n=0}^{+\infty}\frac{\left(-e^{-2\pi \sqrt{f(t)}}\right)^{n}}{\left(n+\frac{1}{2}\right)^{1-k}},
\label{step3}
\end{equation}
and noticing that $\sum_{n=1}^{+\infty}z^{n}/n^{r}=\mbox{Li}_{r}(z)$ is the polylogarithm \cite{jonquiere1889}, which reduces to a polynomial for any integer $r\leq 0$, and that $\sum_{n=0}^{+\infty}z^{n}/(n+a)^{r} =\Phi\left(z,r,a\right)$ is the so-called Lerch trascendent function \cite{lerch1887}, which reduces to a ratio of $y$ polynomials for any $r\leq0$ when $a=1/2$. From Eq. (\ref{step2}) and (\ref{step3}) it follows that
\begin{eqnarray}
\langle\langle\langle x^{2}\rangle\rangle\rangle&=&\frac{L^{2}}{12}+\frac{b(t)}{24}+\frac{c(t)}{2}+\frac{b(t)}{4\pi^{2}}e^{-\frac{L^{2}}{b (t)}}\sum_{m=0}^{+\infty}\left(\frac{L^{2}}{b(t)}\right)^{m}\frac{1}{(2m)!}\sum_{k=0}^{m+1}\frac{(2m+2-k)!}{(m+1-k)!k!}\left[4\pi\sqrt{f(t)}\right]^{k}\mbox{Li}_{2-k}\left(-e^{-2\pi\sqrt{f(t)}}\right)\nonumber \\
&&-\frac{2c(t)}{\pi}e^{-\frac{L^{2}}{b (t)}}\sum_{m=0}^{+\infty}\left(\frac{L^{2}}{b(t)}\right)^{m}\frac{1}{(2m)!}\sum_{k=0}^{m}\frac{(2m-k)!}{(m-k)!k!}\left[4\pi\sqrt{f(t)}\right]^{k} e^{-\pi\sqrt{f(t)}}\Phi\left(-e^{-2\pi\sqrt{f(t)}},1-k,\frac{1}{2}\right).
\label{msdeq3}
\end{eqnarray}
To simplify further the infinite series expression in (\ref{msdeq3}) one can sum the infinite $m$-series for specific values of $k$. For example, if all the $k=0$ and $k=1$ terms in the first series and all the $k=0$ and $k=1$ terms in the second series are summed over all $m$-values, Eq. (\ref{msdeq3}) becomes
\begin{eqnarray}
\langle\langle\langle  x^{2}\rangle\rangle\rangle&=&L^{2}\left[\frac{1}{12}+\frac{\mbox{Li}_{2}\left(-e^{-2\pi\sqrt{f(t)}}\right)}{\pi^{2}}-\frac{2\sqrt{f(t)}}{\pi}\ln\left(1+e^{-2\pi\sqrt{f(t)}}\right)\right]+b(t)\left[\frac{1}{24}+\frac{\mbox{Li}_{2}\left(-e^{-2\pi\sqrt{f(t)}}\right)}{2\pi^{2}}\right.\nonumber \\
&&\left.-\frac{\sqrt{f(t)}}{\pi}\ln\left(1+e^{-2\pi\sqrt{f(t)}}\right)\right]+c(t)\left[\frac{1}{2}-\frac{4}{\pi}\arctan\left(e^{-\pi\sqrt{f(t)}}\right)-4\sqrt{f(t)}\left(1-e^{-\frac{L^{2}}{b(t)}}\right)\frac{e^{-\pi\sqrt{f(t)}}}{1+e^{-2\pi\sqrt{f(t)}}}\right] \nonumber \\
&&+\frac{b(t)}{4\pi^{2}}e^{-\frac{L^{2}}{b (t)}}\sum_{m=1}^{+\infty}\left(\frac{L^{2}}{b(t)}\right)^{m}\frac{1}{(2m)!}\sum_{k=2}^{m+1}\frac{(2m+2-k)!}{(m+1-k)!k!}\left[4\pi\sqrt{f(t)}\right]^{k}\mbox{Li}_{2-k}\left(-e^{-2\pi\sqrt{f(t)}}\right) \nonumber \\
&&-\frac{2c(t)}{\pi}e^{-\frac{L^{2}}{b (t)}}\sum_{m=2}^{+\infty}\left(\frac{L^{2}}{b(t)}\right)^{m}\frac{1}{(2m)!}\sum_{k=2}^{m}\frac{(2m-k)!}{(m-k)!k!}\left[4\pi\sqrt{f(t)}\right]^{k} e^{-\pi\sqrt{f(t)}}\Phi\left(-e^{-2\pi\sqrt{f(t)}},1-k,\frac{1}{2}\right).
\label{msdeqtdc2}
\end{eqnarray}

Continuing this process for all $k$ gives the following single-sum expression, involving the generalized hypergeometric function ${}_2F_2(a_1,a_2;b_1,b_2;z)$ (see e.g. \cite{paris2005} and references therein or \cite{hardy1999} for the properties of the ${}_2F_2$ hypergeomteric function).
\begin{eqnarray}
\langle\langle\langle x^{2}\rangle\rangle\rangle&=&L^{2}\left[\frac{1}{12}+\frac{\mbox{Li}_{2}\left(-e^{-2\pi\sqrt{f(t)}}\right)}{\pi^{2}}-\frac{2\sqrt{f(t)}}{\pi}\ln\left(1+e^{-2\pi\sqrt{f(t)}}\right)\right]+b(t)\left[\frac{1}{24}+\frac{\mbox{Li}_{2}\left(-e^{-2\pi\sqrt{f(t)}}\right)}{2\pi^{2}}\right. \nonumber \\
&&\left.-\frac{\sqrt{f(t)}}{\pi}\ln\left(1+e^{-2\pi\sqrt{f(t)}}\right)\right]+c(t)\left[\frac{1}{2}-\frac{4}{\pi}\arctan\left(e^{-\pi\sqrt{f(t)}}\right)-4\sqrt{f(t)}\left(1-e^{-\frac{L^{2}}{b(t)}}\right)\frac{e^{-\pi\sqrt{f(t)}}}{1+e^{-2\pi\sqrt{f(t)}}}\right] \nonumber \\
&&+\frac{b(t)}{4\pi^2}e^{-L^2/b(t)}\sum_{k=2}^\infty\frac{1}{(2k-2)!}\biggl(\frac{4\pi\sqrt{f(t)}L^2}{b(t)}\biggr)^{k-1}4\pi\sqrt{f(t)}\times \nonumber \\
&&\hspace{5cm}{}_2F_2\Bigl(1+\frac{k}{2},\frac{1}{2}+\frac{k}{2};k,k-\frac{1}{2};\frac{L^2}{b(t)}\Bigr)\mbox{Li}_{2-k}(-e^{-2\pi\sqrt{f(t)}}) \nonumber \\
&&-\frac{2c(t)}{\pi}e^{-L^2/b(t)}\sum_{k=2}^\infty \frac{1}{(2k)!}\biggl(\frac{4\pi\sqrt{f(t)}L^2}{b(t)}\biggr)^{k}e^{-\pi\sqrt{f(t)}}\times \nonumber \\
&&\hspace{5cm}{}_2F_2\Bigl(1+\frac{k}{2},\frac{1}{2}+\frac{k}{2};k+1,k+\frac{1}{2};\frac{L^2}{b(t)}\Bigr)\Phi(-e^{-2\pi\sqrt{f(t)}},1-k,\frac{1}{2})
\end{eqnarray}

\end{document}